\documentclass[sigconf]{acmart}
\usepackage{tabularx}
\usepackage{graphicx}
\usepackage{subcaption}
\usepackage{enumitem}
\usepackage{balance}
\usepackage{mwe}
\AtBeginDocument{%
  \providecommand\BibTeX{{%
    \normalfont B\kern-0.5em{\scshape i\kern-0.25em b}\kern-0.8em\TeX}}}

\acmConference[BiblioDAP'21]{}{}{}

\begin{document}

\title{\emph{Bib2Auth}: Deep Learning Approach for Author Disambiguation using Bibliographic Data}

\author{Zeyd Boukhers \quad Nagaraj Bahubali \quad Abinaya Thulsi Chandrasekaran \quad Adarsh Anand \quad Soniya Manchenahalli Gnanendra Prasad \quad and \quad Sriram Aralappa}



\begin{abstract}
Author name ambiguity remains a critical open problem in digital libraries due to synonymy and homonymy of names. In this paper, we propose a novel approach to link author names to their real-world entities by relying on their co-authorship pattern and area of research. Our supervised deep learning model identifies an author by capturing his/her relationship with his/her co-authors and area of research, which is represented by the titles and sources of the target author's publications. These attributes are encoded by their semantic and symbolic representations. To this end, \emph{Bib2Auth} uses $\sim$ 22K bibliographic records from DBLP 
repository and is trained with each pair of co-authors. The extensive experiments have proved the capability of the approach to distinguish between authors sharing the same name and recognize authors with different name variations. \emph{Bib2Auth} has shown good performance on a relatively large dataset, which qualifies it to be directly integrated into bibliographic indices.
\end{abstract}




\maketitle

\section{Introduction}
\label{introduction}

Ensuring high quality bibliographic data is important for easy and quick access to the literature. One of the most important criteria for this quality is the linking of authors to their real-world entities. Here, an author name denotes a set of character sequences that refer to one or more people~\footnote{It is estimated that about 114 million people share 300 common names.}, whereas real-world author entity indicates a unique author that cannot be identified only by his/her name~\footnote{In the DBLP database, there are 27 exact matches of ‘Chen Li’, 23 reverse matches and more than 1000 partial matches} but with the help of other identifiers such as ORCID. However, in bibliographic data (e.g., references), authors are usually referred to by name only. Given the large number of authors who share the same name (i.e., homonymy), it is difficult to link names in bibliographic sources to their real-world authors, especially when the source of the reference is not available or does not provide indicators of the author's identity. The problem is even more critical when names are substituted by their initials to save space and when they are erroneous due to wrong manual editing as found in our previous work~\cite{zeyd2019end}. Disciplines such as social sciences and humanities suffer more from this problem as most of the publishers are small or medium-sized and cannot ensure a continuous integrity of the bibliographic data.

Table~\ref{tab:illust} demonstrates examples of reference strings covering the above mentioned problems. Here, each author is given an identifier ranging from $\mathrm{a}_\mathrm{1}$ to $\mathrm{a}_\mathrm{7}$. In the first reference string, the author $\mathrm{a}_\mathrm{2}$ can be uniquely identified, whereas $\mathrm{a}_\mathrm{1}$ can refer to any of the four `Bing Li's available in the DBLP network due to \emph{homonymy}. Subsequently, the authors $\mathrm{a}_\mathrm{2}$ and $\mathrm{a}_\mathrm{5}$ refer to the same author `Weihua Xiong' but appear in different name variations due to different citation styles (\emph{synonymy}). Finally, in the third reference string, the author $\mathrm{a}_\mathrm{6}$ refers to `Christophe Ponsard', but the last name is misspelled as `Pinsard' due to wrong manual editing.

 \begin{table}
  \centering
\begin{tabular}{ | m{.4cm} | m{1.5cm}| m{1.1cm} | m{3.9cm} | }
 \hline
 {\bf ID}  & {\bf Issue type} & {\bf Citation styles} & {\bf Citation or reference strings} \\ \hline
1 & Homonymy & ASA & Bing Li($\mathrm{a}_\mathrm{1}$), Weihua
 Xiong($\mathrm{a}_\mathrm{2}$). 2012. Visual saliency map from tensor analysis. In Proceedings of the 26th AAAI Conference on Artificial Intelligence, pages 1585-1591 \\ \hline
2 &   Synonymy  & IEEE &H. Peng($\mathrm{a}_\mathrm{3}$), B. Li($\mathrm{a}_\mathrm{4}$),W. Xiong($\mathrm{a}_\mathrm{5}$), Predicting image memorability by multi-view adaptive regression, Proc. 23rd ACM Int. Conf. Multimedia, pp. 1147-1150, 2015. \\ \hline
3 & Incorrect author name & APA & Pinsard, C.($\mathrm{a}_\mathrm{6}$), Ramdoyal, R.($\mathrm{a}_\mathrm{7}$) : An ocr-enabled digital comic books viewer. In: Computers Helping People with Special Needs, pp. 471–478. Springer, (2012)  \\ \hline 
\end{tabular}
\caption{Illustrative examples of author name ambiguity and incorrect author names}
\label{tab:illust}
\end{table}

To overcome the above mentioned problems, several approaches have been proposed and are integrated in different bibliographic systems~\cite{muller2017semantic,kim2018web,foxcroft2019name2vec, hussain2017survey,ferreira2012brief,torvik2009author,smalheiser2009author,khabsa2014large,khabsa2015online}. Most of these approaches rely on the author name string~\cite{foxcroft2019name2vec}. Other approaches consider the strings of the co-authors and the titles along with the author name strings~\cite{muller2017semantic}. However, none of these papers takes into account different name variations of the same author that occur in the references of different publications.

Author name ambiguity and inaccuracy can affect the quality of scientific data gathering and consequently leads to incorrect identification and credit attribution to authors. In this paper, we address the problem of linking authors occurring in streaming references to their real world-entities. To overcome the inherent challenges, we propose a deep learning model that disambiguates author names by relying on all informative reference elements. The assumption is that any author is most likely to publish articles in specific fields of research. Therefore, we employ articles' titles and sources (i.e. journal, book title, etc.) to bring authors close to their fields of research represented by the titles and sources of publications. This also helps in enhancing the distinction of one author from the other who publish articles in completely different research areas. We also assume that authors who have already published together are more likely to continue collaborating and publish other articles. Therefore, in addition to author names, we train our model on other informative reference elements such
as co-authors.  

For the goal mentioned above, our proposed model \emph{Bib2Auth} is trained on a bibliographic dataset obtained from DBLP, where a sample consists of a pair of co-authors (i.e. target author and a co-author), title, and source (i.e., journal name or book title). For co-authors, the input of \emph{Bib2Auth} is a vector representation obtained by applying Char2Vec. For title and source, the BERT model is used to capture the semantic representations of the sequence of words. \emph{Bib2Auth} has been trained and tested on a very challenging dataset, where a lot of authors share the same names. The main contributions of this paper are: 
\begin{itemize}[leftmargin=*]

\item We proposed a novel approach for author name disambiguation using semantic and symbolic representations.
\item We propose a challenging dataset for author name disambiguation.
\item The experimental results on the challenging dataset demonstrate the effectiveness of our \emph{Bib2Auth} to disambiguate author names.
\end{itemize}

The rest of the paper is organized as follows. Section~\ref{related_work} briefly presents related work. Section~\ref{method} describes the proposed framework. Section~\ref{experiments} presents the dataset, implementation details and the obtained results of the proposed model. Finally, Section~\ref{conclusion} concludes the paper and gives insights on future work.

\section{Related Work}
\label{related_work}

Author name mapping can be considered as one of the cases of the broad notion of entity linkage. Named Entity Linking (NEL), also called Named Entity Disambiguation (NED), is a process of mapping an entity mention to its corresponding unique entity in Knowledge Base (KB) using the surrounding contextual information \cite{singh2018natural}. Various techniques have been used to achieve the task of entity linkage, and these techniques can be broadly categorised into supervised, unsupervised, and graph-based.

\par In the unsupervised setting, Cucerzan in~\cite{cucerzan2007large} studies the computation of semantic relatedness of documents, which is Explicit Semantic Analysis (ESA). Here, the goal is to find a pair of articles that contain similar words and compute their degree of relatedness from the word-based similarity of the articles. Momeni et al.\cite{momeni2016evaluating} present a clustering approach that addresses the problem of author name homonyms, i.e. when different authors share identical names. Here, all publications with certain ambiguous author names are formed into blocks and within each such block, a clustering method is applied to form new clusters representing a unique author entity. To measure the similarity between two publications, co-authors are used. Ferreira et al.\cite{ferreira2010effective} use a similar approach that applies clustering to the citation records of the authors. They propose a hybrid disambiguation method that employs supervised and unsupervised learning. In the unsupervised step, patterns in co-authorship are used to generate clusters of authorship records. Then, in the supervised step, a subset of the filtered clusters is used for training to derive a disambiguation function.

\par In the field of supervised learning, there are only a few works that use traditional approaches, the rest are dominated by deep learning approaches. Han et al.\cite{han2004two} present two supervised learning approaches for disambiguating authors in citations. First, using the Naive Bayes probability model, a generative statistical model to capture all writing patterns in author citations. Second, using Support Vector Machines (SVMs) \cite{vapnik2013nature}, a discriminative model. This paper proposes the idea of a canonical name, a name that is minimally invariant and complete name entity that can be used for disambiguation. Qian et al.\cite{qian2011combining} use High Precision Clusters and High Recall Clusters to train the similarity functions between publications. With these learned similarities, a clustering algorithm is further applied to generate final clusters. Sun et al.\cite{sun2011detecting} manually label 500 authors as ambiguous or not and use the annotated data to further train the model. They use two classes of features for training. The first is a  heuristic based on the percentage of citations collected from the top name variants for an author. The second feature class utilizes crowdsourced data to detect ambiguity at the topic level.

\par In recent years, the field of Deep Learning has been heavily explored by several researchers, often developing models that outperform the more traditional feature-based approaches for Entity Recognition and Linkage. A comprehensive description of the latest neural entity linkage systems is discussed in the survey conducted by Sevgili et al.\cite{sevgili2020neural}.

\par Hourrane et al.\cite{hourrane2018using} propose a corpus-based approach that uses deep learning word embeddings to compute citation similarities. Muhammad Ebraheem et al.\cite{ebraheem2018distributed} propose a novel Entity Resolution system called the DEEPER. It uses a combination of bi-directional recurrent neural networks along with Long Short Term Memory (LSTM) as hidden units to generate a distributed representation (vector) for each tuple to capture the similarities between tuples. Ganguly et al.\cite{ganesh2016author2vec}  propose Author2Vec, a model that uses Deep Learning to overcome the link sparsity problem of models such as DeepWalk \cite{liben2007link}. Author2Vec uses dedicated models for content (text data) and link information, which are used in parallel. By combining these two models, Author2Vec intends to learn author embeddings for a bibliographic network in an unsupervised manner. Raiman et al.\cite{raiman2018deeptype} propose DeepType, a multilingual Entity Linking from Neural Type System Evolution. It has been applied on the datasets WikiDisamb30, CoNLL (YAGO), TAC KBP 2010) with very encouraging results.\\

\par Graph-based methods consider the citation network as a graph where each node can be one of the citation attributes such as author, title, publication venue, etc. The edges in such graphs represent the relationship between any two citation attributes. Han et al.\cite{han2011collective} propose a graph-based method where
the global interdependence between the entity linking decisions is modeled as a graph and a purely collective
algorithm is further applied to disambiguate the name mentions of entities in the document. Hoffart et al.\cite{hoffart2011robust}   present a robust method for collective disambiguation of author names. It harnesses the context from a knowledge base and uses a new form of coherence graph. The system generates a weighted graph of candidate entities and mentions computing a dense subgraph that approximates the best entity-mention mapping.
\section{\emph{Bib2Auth}}
\label{method}

\begin{figure*}
  \centering
  \includegraphics[width=5in,height=2.5in]{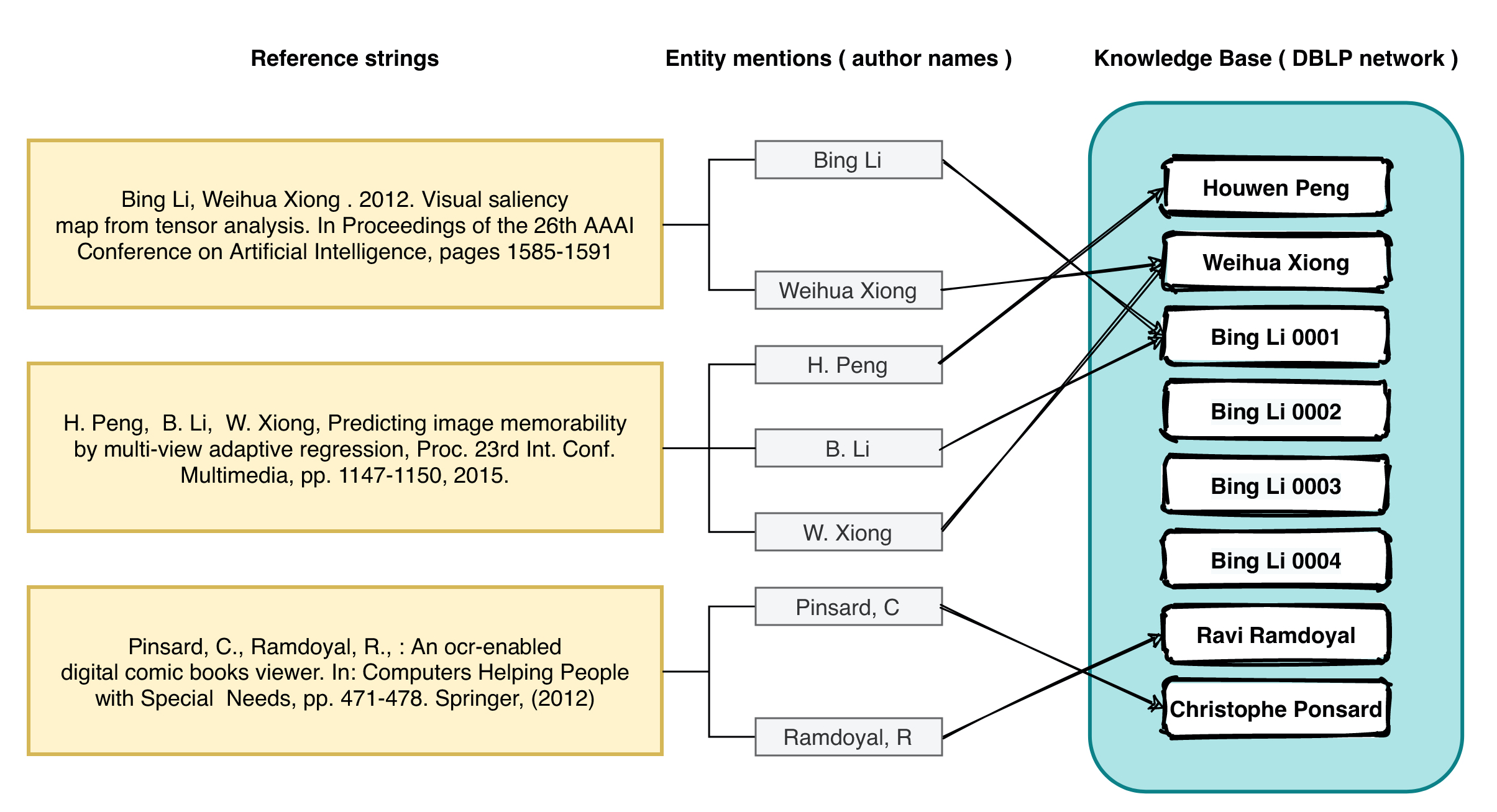}
  \caption{An illustration for the task of linking an name mention in the reference string with the corresponding DBLP author entity.}
  \label{fig:illust}
\end{figure*}

In this paper, the process of author name disambiguation is defined formally as: Let $R=\{r_1, r_2, \cdots, r_M \}$ be a set of $M$ reference strings, $A=\{a_1, a_2, \cdots, a_N \}$ be a set of $N$ author name mentions and $E=\{e_1, e_2, \cdots, e_K \}$ be a set of $K$ unique author entities in DBLP. The goal of \emph{Bib2Auth} is to map each name mention (i.e. author name) $a_n \in A$ in the reference $r_m \in R$ to the corresponding unique entity $e_k \in E$ with DBLP identifier. Figure~\ref{fig:illust} illustrates this task of author name mapping.\\

For this task, we propose \emph{Bib2Auth}, a supervised deep learning model that maps each author name in the reference string to its corresponding unique author present in the DBLP database. This is achieved with the assistance of the contextual citation attributes such as co-author names, journal name (or book title), and title of the paper. We transform each of these contextual attributes into the corresponding embeddings depending on their type. Further, these embeddings are collated and supplied as input to the model.  

\subsection{Model Architecture}

\emph{Bib2Auth} is a Neural Network model with two input layers. The first input layer represents the concatenation of the co-author embedding and the content embedding. The latter is the average embedding of title and source. The second input layer represents the embedding of the target author. For author and co-author, the embedding is of length $200$ and is generated using Char2Vec~\cite{cao2016joint}, which is a recurrent neural network that provides a symbolic representation of the given word. Char2Vec is a powerful model as it can put two words with a slight difference very close to each other in the representation space. For title and source, the embedding is of length $786$ and is generated using BERT~\cite{devlin2018bert} which provides a vector representation of words w.r.t their context in the sentence. The goal of separating the two inputs is to overcome the sparseness of the content embedding and force the model to emphasise more on target author representation.

\begin{figure*}
  \centering
  \includegraphics[height=2.5in]{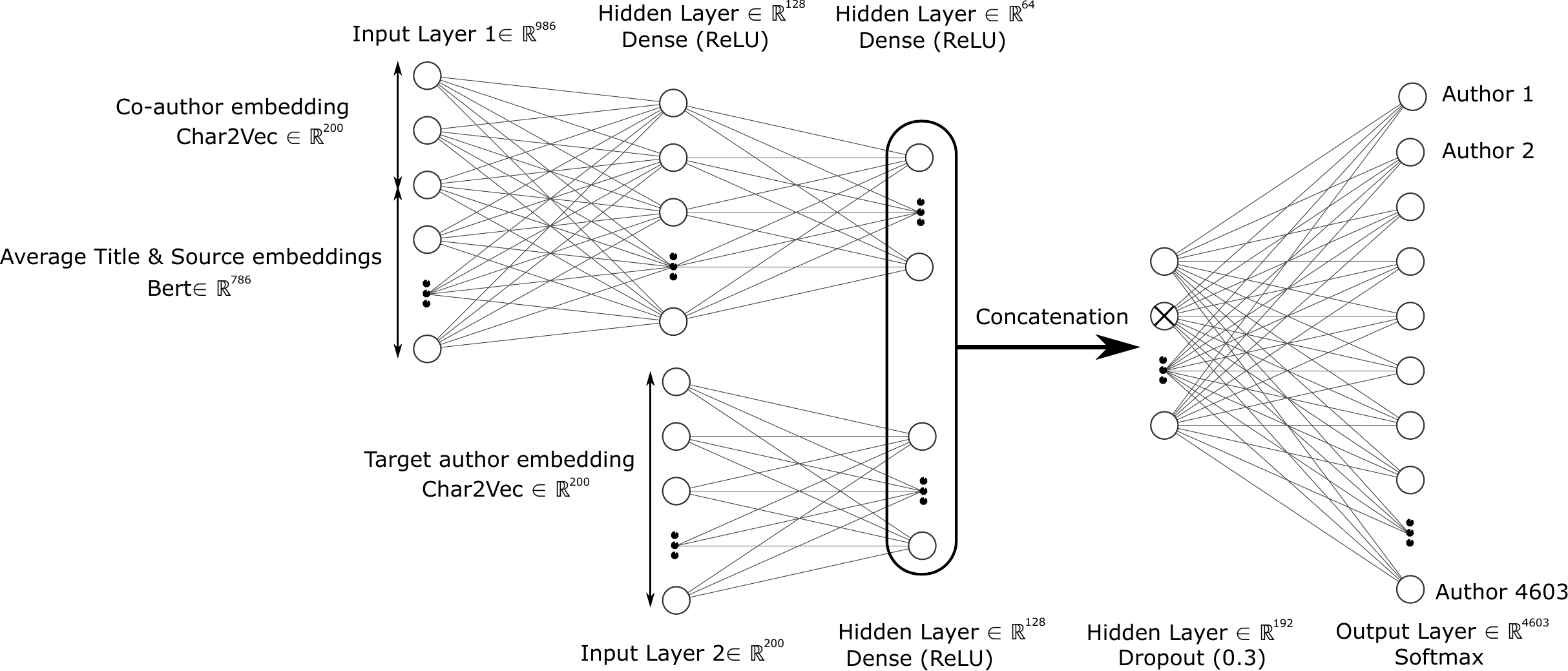}
  \caption{The architecture of \emph{Bib2Auth} model.}
  \label{fig:arch}
\end{figure*}

Figure~\ref{fig:arch} illustrates the architecture of \emph{Bib2Auth} model, with an output layer of length 4603, which corresponds to the number of authors in our dataset. All hidden layers possess ReLU activation function, while the output layer possesses a Softmax activation function. Since the model needs to classify thousands of classes, each of which is represented with very few samples, $30\%$ of the units in the last hidden layer are dropped out during training to avoid overfitting. Furthermore, the number of publications differs significantly from one author to another. Therefore, each class (i.e., author) is weighted according to the number of its samples (i.e., publications). The model is trained using \emph{adam} optimizer and the sparse categorical cross-entropy loss function.

\subsection{Author name embedding}
Author names do not hold any specific semantic nature as they are simply a specific sequence of characters referring to one or more persons. Besides, language models such as Word2vec or GloVe essentially work on a finite set of vocabulary and fail to interpret words that are out of vocabulary. We require a model that can encode the words based on the order and distribution of characters such that authors with similar name spellings are encoded similarly.

Chars2vec is a language model that is preferred when the text consists of abbreviations, typos, etc. It captures the non - vocabulary words and places words with similar spellings closer in the vector space. This model uses a fixed list of characters for word vectorization, where each character is represented by a one-hot encoding.

A pair of similar or dissimilar words is passed to this model,  where the sequence of one-hot vectors of each letter in the words is passed through two layers of LSTM and to an extended neural network. This model strives to place similar words nearer in the vector space by reducing the distance between words in similar pairs. In contrast, it tries to increase the distance between words in dissimilar pairs and place them far apart.

To obtain the embedding of author names, we have used the Chars2vec model, which is implemented using Keras based on TensorFlow. Our aim here is to place the different citation styles of an author close to each other in the vector space.

\subsection{Source embedding}

Journal name and book title can provide a hint about the area of research or domain to which published papers relate. Subsequently, this information can be used to enhance the encoding of an author to reflect his area of research. In the DBLP network, we noticed that a significant number of reference strings contain a journal name that is abbreviated. To ensure that the essence of journal names is captured, we fetched the expanded journal names using the DBLP Venue API\footnote{\url{https://dblp.org/search/venue/api}}. The Venue API returns data in an HTML format upon which we used the beautiful soup\footnote{\url{https://www.crummy.com/software/BeautifulSoup/}} to retrieve the journal names. Currently, the total number of unique journals that exist in the DBLP network is $13,101$. Once we have the journal names and book titles, the sentence embeddings are generated using the pre-trained BERT model~\cite{devlin2018bert} 

\subsection{Title embedding}

Titles of references are always meaningful sentences and infer the domain of the scientific paper. This feature is also used to understand the research area of the author. Similar to source, we used pre-trained BERT model to generate sentence embedding for the titles.

\subsection{Model Tuning}

For each training epoch, \emph{Bib2Auth} model fine-tunes the parameters to predict the appropriate target author. The performance of the model is significantly affected by the number of epochs set for training. In particular, a low epoch number can lead to underfitting. A high epoch number, on the other hand, can lead to overfitting. To avoid this, we have enabled early stopping, which allows the model to set an arbitrarily large number for the epoch.

Keras supports early stopping of the training via a callback called \emph{EarlyStopping}. This callback is configured using \emph{monitor} argument, which can be used to monitor the validation loss. With this setup, the model receives a trigger to halt the training when it no longer observes improvement in validation loss.

Often, the very first indication that the validation loss is no longer improving is not the correct epoch to stop training, as the model may improve again after passing through a few more epochs. We overcome this by adding a delay to the trigger in the form of the number of consecutive epochs we can wait for to observe no further improvement.  A delay is added by setting the \emph{patience} argument to an appropriate value. \emph{Patience} in \emph{Bib2Auth} is set to $50$, so that the model stops training only if the validation loss does not improve in the last 50 consecutive epochs.

\subsection{Model checkpoint}
Although \emph{Bib2Auth} stops the training process when it reaches a minimum validation loss, the model obtained at the end of the training may not give the best accuracy on validation data. To account for this, Keras provides an additional callback called \emph{ModelCheckpoint}. This callback is configured using another \emph{monitor} argument. We set the \emph{monitor} to monitor the validation accuracy. With this setup, the model updates the weights only when it observes better validation accuracy compared to earlier epochs. Eventually, we end up persisting the best state of the model with respect to the best validation accuracy.
\section{Experiments}
\label{experiments}

In this section, we present the experimental validation of \emph{Bib2Auth} on the dataset collected from DBLP database. 

\subsection{Dataset}
\begin{figure}[h!]
  \includegraphics[width=0.5\textwidth]{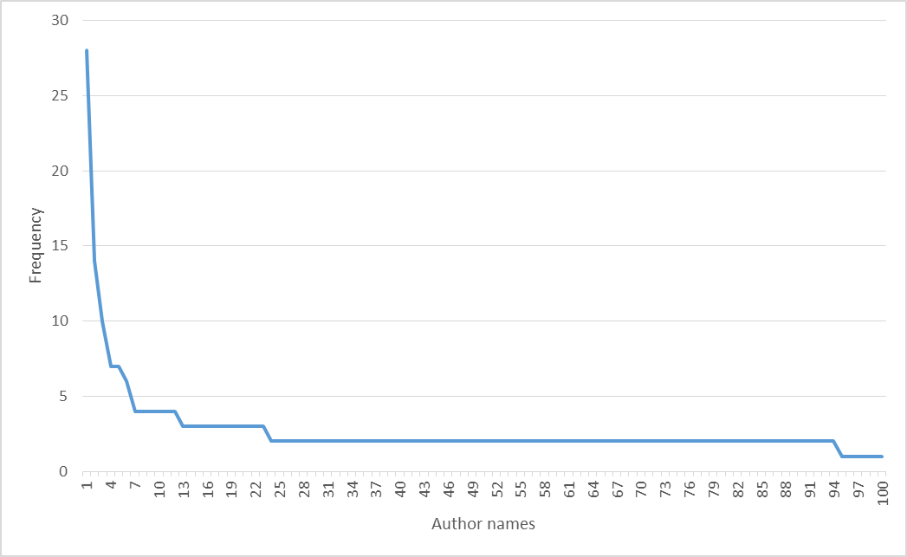}
  \caption{The frequency of author names in the used dataset.}
  \label{fig:freq}
\end{figure}

We have used the DBLP bibliographic repository\footnote{\url{https://dblp.uni-trier.de/xml/}} as a source of data to train, validate and test \emph{Bib2Auth}. The DBLP version of July 2020 contains 4.4 million bibliographic records such as conference papers, articles, thesis, etc., from various fields of research. Each record represents metadata information of a publication with one or more authors, title, journal, year of publication and a few other attributes. The number of these attributes differs from one reference to another. Also, the authors in DBLP who share the same name have a suffix number to differentiate them. For instance, the authors with the same name 'Bing Li' are given suffixes such as 'Bing Li 0001', 'Bing Li 0002'. Figure~\ref{fig:freq} shows the frequency of the first 100 authors sharing the same name (first and last names) in the dataset used by \emph{Bib2Auth}.

Since training the model on 4.4 million records is very time-consuming, we used a sample dataset of $21802$ records which are randomly sampled from the DBLP repository. However, we ensured that the dataset contains records from different authors sharing the same first and last name. Since authors publish papers with different co-authors, it is challenging to sample a dataset while ensuring that all or at least some publications from all co-authors are present in the training set. Otherwise, the sampling process will end up retrieving all records. Therefore, the sampling process starts by selecting a finite set of records, where all co-authors are present with around $70\%$ of their publications. This led to sampling of $2534$ records with $4603$ unique co-authors for training.

\begin{figure}[h!]
  \includegraphics[width=0.5\textwidth]{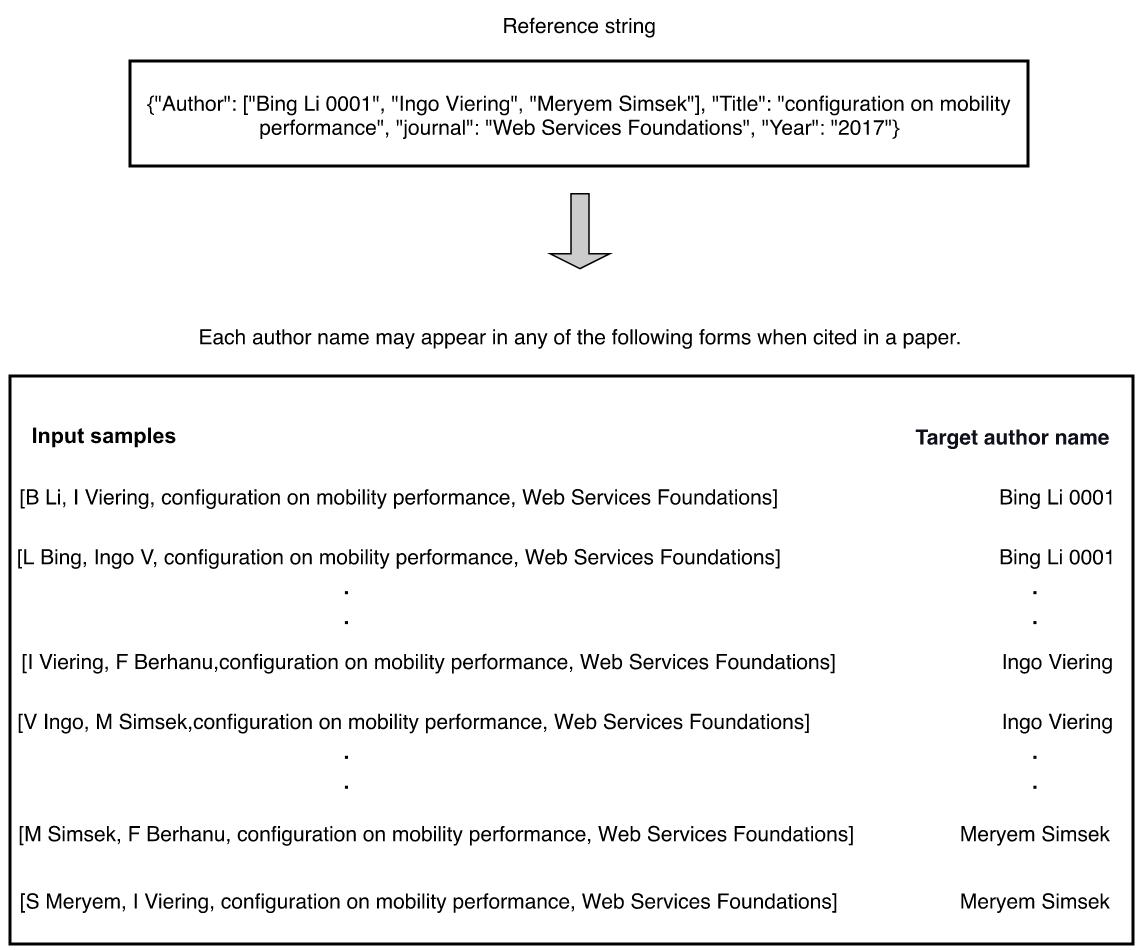}
  \caption{An example of generating input samples.}
  \label{fig:exmp}
\end{figure}

As each record may have one or more authors, it is not possible to feed input samples of variable length to our model. Therefore, we limit the number of authors in a sample to two. One acts as a target author and the other as a co-author.  Thus, we generate all possible name combinations of author-co-author pairs from each record. This means that each record is represented by multiple input samples with the same title and source, but with different pairs of author names. Note that all co-authors are considered as target authors in the generation of input samples. To account for possible citation styles for author names, we further enrich the input sample by including all name variations for author and co-author. If the first and last name of the target author and co-author consist of one token, six pairs of variations are generated. Consequently, each sample is a tuple consisting of an author name variant, a co-author name variant, a title, and a journal. Figure~\ref{fig:exmp} illustrates an example of generating multiple samples from one bibliographic record to capture possible name variants of authors. With this strategy, we assume that the model can capture all possible variations of author names. The class label of each such sample is the original author from the DBLP corresponding to the author name variant.

Subsequently, for each author in the training set, we sample new records for validation and testing, by trying to ensure that these records contain at least $2$ co-authors from the training set. Note that for these records, only combinations of co-authors that were present in the training process are considered for validation and testing. The reason for preparing the data in such a way is to ensure that records from training are not used either in validation or testing. Specifically, each of the three sets has unique publications (i.e., titles). The statistical details of the data split are shown in Table~\ref{tab:data}.

\begin{table}[]
    \centering
    \begin{tabular}{|c|c|c|c|}
    \hline
         & Combination & Records & \# unique authors \\
         \hline
       Training set  & $45060 (\sim 66\%)$ & $2534$ & $|C|=4603$\\
       \hline
       Validation set & $11584 (\sim 17\%)$ & $10952$ & $2537 \in C$\\
       \hline
       Testing set & $11584 (\sim 17\%)$ & $10850$ &  $2576 \in C$\\
       \hline
    \end{tabular}
    \caption{Statistical details of the used dataset.}
    \label{tab:data}
\end{table}

\subsection{Results}
To our knowledge, there is no approach that addresses the problem of author name disambiguation with a variety of names. Besides, all these approaches use different datasets with a big challenge disparity without providing the datasets or the source code~\cite{hussain2017survey}. Therefore, it is not possible to compare our \emph{Bib2Auth} against baseline approaches. Our source code and the used dataset are publicly available\footnote{\url{https://github.com/orgs/ZResearch/teams/bib2auth}}.

We evaluated our model on train, validation and test data with a split ratio of \text{66\%} : \text{17\%} : \text{17\%}. The training and validation loss is monitored continuously over the optimal number of epochs. We run the model on each combination of the test set. Table~\ref{tab:result1} represents the Macro and Micro average of precision, recall, and F1-score.

\begin{table}[]
    \centering
    \begin{tabular}{|c|c|c|}
    \hline
         &  Macro & Micro\\
         \hline
        Precision & $0.989$ & $0.975$\\
        \hline
        Recall & $0991$ & $0.975$\\
        \hline
        F1-Score & $988$ & $0.975$\\
        \hline
    \end{tabular}
    \caption{The obtained results of \emph{Bib2Auth} on test dataset for each pair of co-authors.}
    \label{tab:result1}
\end{table}

The results of Table \ref{tab:result1} are based on the individual classification of each pair of co-authors. However, in real scenarios, all authors are available, which can increase the accuracy of the classification. Unfortunately, we were not able to provide such a result as in the test data most of the bibliographic records are presented by two co-authors. As mentioned before, if we consider all authors, they have to be present in the training set as well, which means that all records in the DBLP repository have to be considered.

Moreover, Table~\ref{tab:result2} shows the obtained results of the same model on only one name variant corresponding to the full name of the authors. Table~\ref{tab:result1} and Table~\ref{tab:result2} demonstrates the ability of \emph{Bib2Auth} to distinguish between authors sharing the same name when only a co-author, title, and source of publication are given. The obtained results also prove the effectiveness of the model in handling different name variants. The model can be tested online at the following link\footnote{\url{http://zeyd.boukhers/research/bib2auth}}.

\begin{table}[]
    \centering
    \begin{tabular}{|c|c|c|}
    \hline
         &  Macro & Micro\\
         \hline
        Precision & $0.993$ & $0.984$\\
        \hline
        Recall & $0996$ & $0.984$\\
        \hline
        F1-Score & $994$ & $0.984$\\
        \hline
    \end{tabular}
    \caption{The obtained results of \emph{Bib2Auth} on test dataset for each pair of co-authors, considering only full names.}
    \label{tab:result2}
\end{table}

\subsection{Limitations and obstacles of \emph{Bib2Auth}}
Although \emph{Bib2Auth} has shown promising and satisfactory results, it has several limitations and faces some obstacles, similar to all available methods. In the following, we list the limitations of \emph{Bib2Auth} that we aim to overcome in future work: 

\begin{itemize}[leftmargin=*]
    \item \emph{Bib2Auth} handles new authors (not seen in the training) by setting a confidence threshold. If the classification output is lower than the threshold, \emph{Bib2Auth} assumes that the author is new.
    
    \item When two authors appear together for the first time in a new reference, the model does not completely benefit from this co-authorship because it is not present in the training set.  
    \textbf{Our planned solution:}  We will train an independent model to embed the author's discipline using his/her known publications.  Here, we assume that authors working in the same area of research are close to each other and even if they have not published a paper together, the model would be able to capture the potential co-authorship between a pair of authors. However, training these models is extremely expensive.  
    
    \item So far, \emph{Bib2Auth} considers a small part of the DBLP repository. Training \emph{Bib2Auth} on the entire dataset is extremely expensive.

    \item The accuracy of the DBLP repository is not guaranteed, as mentioned by the maintainers of the platform~\footnote{\url{https://dblp.org/faq/How+accurate+is+the+data+in+dblp.html}}. 
    
    \item An author with a common name might extend his/her research work by co-authoring a new paper in a different research area that he/she used to work on. This leads to different semantical title words. This makes \emph{Bib2Auth} harder to predict when a title with a different set of words is encountered.  
    \textbf{Our planned solution:} We plan to develop an efficient method to determine the author's areas of research interest by mining domain-specific keywords from the entire paper instead of the title, assuming that the author uses similar keywords even in different domains of research. 
    
\end{itemize}

\section{Conclusion}
\label{conclusion}

In this paper, we briefly outlined the challenges associated with entity linking for author names in citation strings. For this, we proposed \emph{Bib2Auth}, a novel framework for author name disambiguation. \emph{Bib2Auth} is a supervised deep learning model which leverages different reference attributes such as co-authors, title, and journal to perform collective disambiguation. Experimental results have shown that \emph{Bib2Auth} achieves very promising and satisfactory results on a challenging dataset, which is full of authors sharing the same names.\par

Despite achieving high accuracy, there are several avenues for improving and enhancing \emph{Bib2Auth}. First, we plan to train the model on the entire DBLP repository by following a hierarchical classification so that authors with completely dissimilar names can be pre-filtered before training. This allows training on a large number of bibliographic records as well as on a high number of authors. Consequently, the time required for training and testing will be reduced.

\section*{Venue}
This paper has been accepted and presented at the non-archival workshop: BiblioDAP@KDD2021~\footnote{\url{https:\\bibliodap.uni-koblenz.de}}~\cite{boukhers2021bibliodap}.
\balance
\bibliographystyle{ACM-Reference-Format}
\bibliography{acmart.bib}

\end{document}